\documentclass[journal]{IEEEtran}
\usepackage{cite}
\usepackage{amsmath,amssymb,amsfonts}
\usepackage{algorithmic}
\usepackage{graphicx}
\usepackage{url}
\usepackage{textcomp}
\newcommand{\ra}[1]{\renewcommand{\arraystretch}{#1}}
\usepackage{booktabs}
\usepackage{xcolor}

\def\BibTeX{{\rm B\kern-.05em{\sc i\kern-.025em b}\kern-.08em
    T\kern-.1667em\lower.7ex\hbox{E}\kern-.125emX}}

\usepackage{todonotes}

\title{Modeling Interconnected Social and Technical Risks in Open Source Software Ecosystems}

\author{William Schueller and Johannes Wachs
\thanks{W. Schueller is with the Complexity Science Hub Vienna and the Medical University of Vienna. J. Wachs is with the Vienna University of Economics and Business and the Complexity Science Hub Vienna. Email: johannes.wachs@wu.ac.at}
}

\begin{document}

\maketitle

\begin{abstract}
Open source software ecosystems consist of thousands of interdependent libraries, which users can combine to great effect. Recent work has pointed out two kinds of risks in these systems: that technical problems like bugs and vulnerabilities can spread through dependency links, and that relatively few developers are responsible for maintaining even the most widely used libraries. However, a more holistic diagnosis of systemic risk in software ecosystem should consider how these social and technical sources of risk interact and amplify one another. Motivated by the observation that the same individuals maintain several libraries within dependency networks, we present a methodological framework to measure risk in software ecosystems as a function of both dependencies and developers. In our models, a library's chance of failure increases as its developers leave and as its upstream dependencies fail. We apply our method to data from the Rust ecosystem, highlighting several systemically important libraries that are overlooked when only considering technical dependencies. We compare potential interventions, seeking better ways to deploy limited developer resources with a view to improving overall ecosystem health and software supply chain resilience.

\end{abstract}



\section{Introduction}

Open source software (OSS) ecosystems are built by decentralized collaborations of thousands of software developers. Relative to the number of developers active in the system as a whole, individual libraries are maintained by small teams or even individuals. The dependencies between these libraries form the backbone of the ecosystem, allowing developers to focus on highly specialized work by importing the work of others. The resulting dependency network represents both a valuable distribution of work and effort across developers, and a potential source of risk. Errors, bugs, and vulnerabilities can propagate through this network \cite{decan2019empirical,eghbal2020working} and we rely on the maintainers of the libraries we use to address these issues \cite{avelino2016novel}. Yet a more complete conceptualization of systemic risk in OSS ecosystems should recognize that these are correlated risks in an interconnected system: key individuals maintain multiple libraries up and down the dependency network.

Indeed recent events have shown how a variety of issues can propagate through software dependency networks, causing outages and problems in important real-world systems. The growth of dependencies in multiple ecosystems \cite{decan2019empirical} and the proliferation of vulnerabilities through those dependencies \cite{decan2018impact} are well-documented \cite{ohm2020backstabber}. Another line of research, highlighting a different aspect of OSS ecosystem fragility, focuses on the developers maintaining these libraries. These works examine the social and structural patterns associated with collaboration and sustained activity in OSS \cite{qiu2019going,valiev2018ecosystem}. One often replicated finding is that many of the most important libraries are maintained by small groups or individuals \cite{avelino2016novel,pfeiffer2021identifying}.

\begin{figure*}
    \centering
    \includegraphics[width=1.02\textwidth]{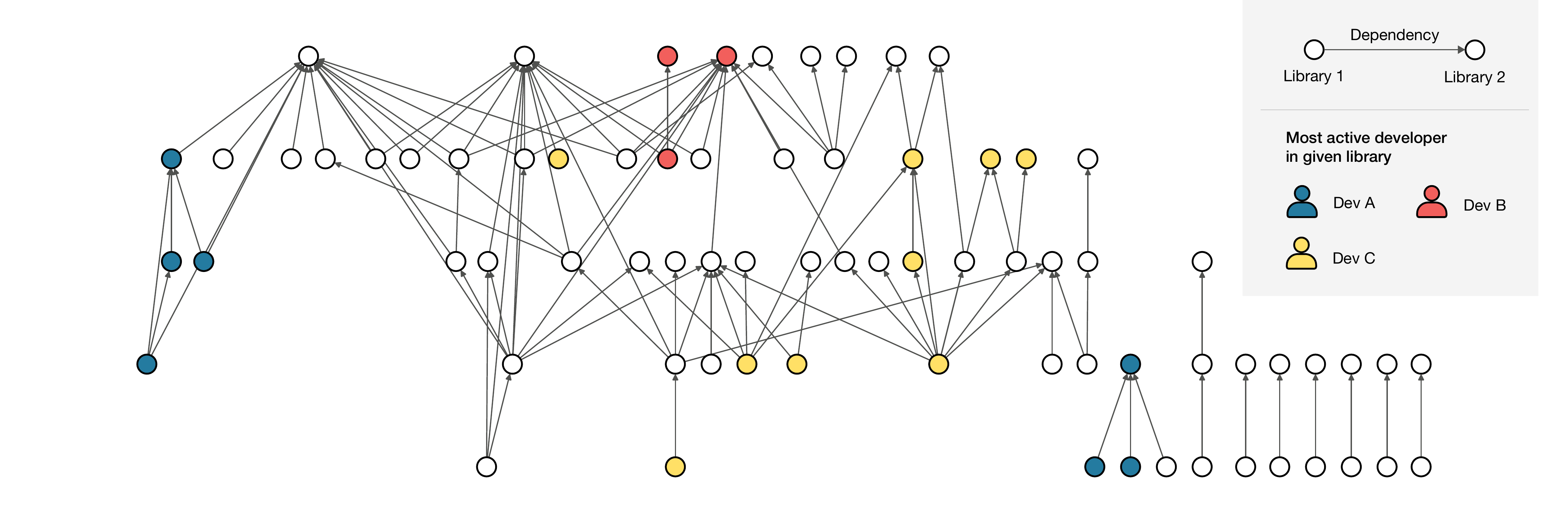}
    \caption{The dependency network among the 100 most downloaded libraries in the Rust ecosystem, observed July 2020. A directed edge between two libraries indicates a dependency. Distinct colors highlight three groups of libraries that have the same developer making the most commits in the previous year. For example, Developer B makes the most commits in each of the three red libraries. Across all 100 libraries, there are 59 unique most active developers, indicating that key individuals often play an important role in multiple interdependent libraries. Note: 24 disconnected libraries are not shown.}
    \label{fig:motiviating_figure}
\end{figure*}

While these two perspectives certainly each highlight important problems on their own, we argue that significant systemic risks in OSS ecosystem emerge through the complex \textit{interaction} of their social and technical systems. The risks of ever-expanding dependency networks are amplified when individuals are core contributors to several libraries in a dependency chain. As we can observe in Figure \ref{fig:motiviating_figure}, in which we plot the dependencies among the 100 most downloaded libraries in the Rust ecosystem, it often happens that individual developers are the most prolific contributors to multiple libraries. The departure of any of these developers would introduce a correlated shock affecting the future functionality of several libraries in the system. The aim of our work is to establish a framework to quantify the systemic importance of \textit{all} developers and libraries that takes these potential correlations into account.

In particular, we adapt methods used to study the propagation of errors in complex systems to the case of open source software ecosystems. We use a simulation approach to quantify systemic risk and apply it to data from the Rust ecosystem. In particular, we simulate the removal of developers from the system, which induces potential failures in libraries that they maintain, which in turn spread with some probability to downstream dependencies. The likelihood that a library fails, governed by a \textit{production function}, increases in the share of its developers who have left and as its upstreams fail. We quantify this likelihood using production functions, modeling maintainers and functioning upstream dependencies as playing a complementary role in insuring the survival of a library. We define an iterative equation to calculate the spread of issues resulting from the departure of specific developers from the system. At the system level we find the significant potential for long cascades of failures when specific individuals leave. We also highlight the libraries that play a major role in many potential cascade paths - which present natural intervention points.

Among the top 1,000 Rust libraries by count of their downstream dependencies, our measure of an individual library's systemic importance is moderately correlated with how many direct (Spearman's $\rho \approx .56$) and transitive dependencies (Spearman's $\rho \approx .54$) it has, and thus highlights key libraries that these purely technical dependency-based measures overlook. Our measure has an even weaker correlation with the number of GitHub stars a library has (Spearman's $\rho \approx .42$), suggesting that social visibility is a poor proxy for systemic importance. This motivated us to use our framework to evaluate potential interventions, i.e. by allocating developer resources to libraries according to various heuristics. We show that allocating developers to the most systemically risky libraries improves system robustness more than by adding developers to libraries central in the dependency network or to socially popular libraries. We argue that this provides valuable insights for individuals, foundations, and firms who are seeking to support OSS ecosystem stability by sponsoring developers or by contributing directly \cite{spaeth2015research,overney2020not}.

We now proceed with a review of related works and then introduce the Rust ecosystem and the data we use. We then describe the simulations we carry out and report results. We consider various alternative intervention strategies aiming to improve the global health of an ecosystem by deploying developers to key libraries. The paper concludes with a discussion of limitations and ideas for future work.

\section{Background}

In this section we review related work, first on the measurement of systemic risks and cascades in complex systems. We then turn to the specific case of such risks in software ecosystems, discussing both social and technical factors.

\subsection{Resilience and Vulnerability of Complex Systems}
The field of complexity science has long studied the vulnerability of interconnected systems. Studies of cascading failures in financial networks \cite{haldane2011systemic,thurner2013debtrank}, supply chains \cite{diem2021quantifying}, power distribution networks \cite{kinney2005modeling}, regional economies \cite{toth2022technology}, and healthcare systems \cite{sardo2019quantification} all highlight that a few key nodes in a system can play a systemically important role that is not obvious from their local topology or individual size. The spread of errors in coupled networks, and thus the identity of the most important nodes vis-a-vis such spreading, is even less predictable~ \cite{schneider2013towards,poledna2015multi}. Previous work in the software engineering research community has pointed out the potential application of complex systems approaches to the study of software systems~\cite{mens2015ecology,decan2019empirical}. However, to our knowledge most work in this area so far has been descriptive.

These studies of complex systems simulate the spread of failures from specific sources to quantify overall systemic risk and the importance of individual entities in the system ~\cite{peters2008modelling}. In these simulations, the functions governing how errors spread are tailored to the specific situation. For instance, recent work on the resilience of supply chains has used production functions such as the Cobb-Douglas and Leontief functions to model the effect of upstream failures on a node's production~\cite{diem2021quantifying}. The choice of a specific production function allows us to consider whether inputs are complementary or whether they can be substituted for each other. In the case of the production and maintenance of software libraries, the effort of maintainers and the functionality of upstream dependencies are complementary, in general they cannot replace one another. The initial condition of the error or failure is also important: for instance contagion in financial networks often begins with the default of a large loan or a bankruptcy. In the case of any specific application, care must be taken to understand conceptually the mechanisms of how errors spread. 

\subsection{Vulnerabilities of OSS Ecosystems}

\subsubsection*{Technical Vulnerabilities}

Bugs and vulnerabilities spread through software ecosystems via dependencies \cite{valiev2018ecosystem}. A 2018 study estimates that half of libraries in the NPM ecosystem are affected by upstream vulnerabilities \cite{decan2018impact}. A series of real world examples highlight the multiple ways problems can spread through and affect whole software ecosystems via their dependency networks. In 2016 a developer of an auxiliary string formatting program called \textit{left-pad} removed his libraries from NPM, a package manager for JavaScript libraries, and caused a cascade of failures that lead to large scale service interruptions around the web~\cite{hejderup2018software}. A significant share of the web's infrastructure depended, often indirectly, on \textit{left-pad's} 11 lines of code. 

While in this case a reasonable solution would be for OSS developers to remove the dependency and implement their own version of this short script, upstream issues sometimes occur in more substantial pieces of software.  Many of these systemically important libraries are overlooked because they have worked well in the background for many years. The Heartbleed bug, introduced into the widely used OpenSSL cryptography library in 2012, made roughly half a million webservers and their user passwords and cookies vulnerable to attack \cite{durumeric2014matter}. Though the bug was quickly resolved, servers remained vulnerable until patched. 

Software systems are also frequently attacked via loopholes introduced by upstream dependencies \cite{ohm2020backstabber}.  In 2017, intruders exploited a vulnerability to access an Equifax database, exposing personal finance data of over one hundred million people. Equifax used an outdated version of Apache Struts 2, an OS web application framework which had a publicly-known (and patched) security vulnerability \cite{luszcz2018apache}. Another example of software that is widely relied upon is \textit{log4j}, a ``ubiquitous'' Java logging library used widely in enterprise software \cite{lily2021fire}. In late 2021, a zero-day vulnerability in \textit{log4j} was reported which can be used to take control of software systems remotely. In other words, software using \textit{log4j} became vulnerable. Security researchers have recently demonstrated how dependency managers themselves can be used to introduce malicious lines into the codebases of leading software companies \cite{birsan2021hack}. A recent study suggests that the share of libraries in NPM inheriting vulnerabilities from upstream dependencies is rising over time \cite{zerouali2021impact}. Tracking vulnerabilities is also a significant challenge for developers and companies \cite{pashchenko2018vulnerable}, and they often persist in ecosystems \cite{alfadel2021empirical}, for instance when patches are not adopted by downstream dependencies \cite{decan2021back}. 

These risks in the ``supply chain'' of OSS are increasingly recognized and quantified in the empirical software engineering literature \cite{ma2018constructing,amreen2019methodology}. Decan, Mens, and Grosjean describe the evolution of the dependency networks of seven large OSS ecosystems, finding an increasing trend in the number of direct and indirect (sometimes called transitive) dependencies in all of them \cite{decan2019empirical}. Many of these upstream dependencies are small libraries in the mold of \textit{left-pad} \cite{abdalkareem2017developers}. These so called ``trivial'' libraries are ironically more likely to occupy critical positions in the dependency network, owing to their widespread use \cite{chowdhury2021untriviality}. This highlights the importance of considering the network as a whole, rather than focusing on important seeming libraries. Indeed, the relative position of a library in a dependency network is a strong predictor of continued activity \cite{valiev2018ecosystem}. In particular, having dormant upstreams increases the risk that a library will itself become dormant.

\subsubsection*{Social roots of ecosystem vulnerability}
Early advocates for the OSS model of software development argued that small contributions of many developers would lead to high quality software \cite{raymond1999cathedral}. And although the decentralized peer production process has resulted in remarkably successful software \cite{benkler2015peer}, the reliance on volunteers and unpaid labor to maintain such widely used software has led to an \textit{underproduction} of OSS. In this context underproduction, coined by Champion and Mako Hill, refers to a mismatch between the supply of software development labor and demand of people relying on a particular software library \cite{champion2021underproduction}.  

The extent to which software relies on individual developers has been conceptualized as the \textit{truck factor} or \textit{bus factor} of a library \cite{williams2003pair,torchiano2011my}. It has been described as ``the number of developers on a team who have to be hit with a truck (i.e., to go on vacation, to become ill, or to leave the company for another) before the project is in serious trouble'' (see: \url{http://www.agileadvice.com/archives/2005/05/truck}). In other words, the truck factor describes the distribution and redundancy of essential knowledge and know-how about a specific software library or project among its developers. A hypothetical library with a truck factor of one relies in some essential way on the contributions, efforts, and knowledge of a single individual. Empirical studies have shown that truck factors of even very widely used libraries are often very low \cite{ferreira2019algorithms,pfeiffer2021identifying}. One study of 133 popular projects on GitHub found that nearly two-thirds had a truck factor of two or less \cite{avelino2016novel}. In practice, libraries are often abandoned because their original core developers and maintainers lack the time or interest to continue working on them \cite{coelho2017modern}. On the other hand, libraries that have contributors that contribute to other libraries tend to stay active \cite{valiev2018ecosystem}.

When libraries go under-maintained or become deprecated, issues tend to build up. It is one of Lehman's Laws that software quickly becomes ineffective or nonfunctional without maintenance \cite{lehman1980programs}. In practice, the same individuals who write the original code of a program are the ones who maintain it, a task which often requires quick interventions when something goes wrong \cite{cook2020above}. Unmaintained libraries are not adapted to changes of the broader ecosystem. When upstream libraries introduce breaking changes, a deprecated library will cease to function properly and pass issues downstream. This is not just a theoretical concern: over half of NPM libraries depend transitively on at least one deprecated library \cite{cogo2021deprecation}. 

Previous work diagnosing the health of ecosystems has not directly addressed the phenomenon of developers working on multiple libraries within an ecosystem. Such developers can make highly valuable contributions, for example because they facilitate coordination and communication between interlinking parts of a larger system \cite{herbsleb1999splitting} or because they are uniquely placed to anticipate failures or issues \cite{cataldo2012coordination}. Although their attention may be divided \cite{vasilescu2016sky}, developers involved in multiple part of an ecosystem are in a position to better consider how new developments in one library may affect others. At the same time, these same aspects make such developers essential to the system as a whole. When such a developer leaves the OSS world, whether it is because they find a new job and no longer have the time, or because they retire, or simply because they no longer want to participate, they may leave several key libraries under- or unmaintained at the same time.

\section{Data}
We now turn to the data we use to build and test our systemic risk measurement framework. We use data from the Rust ecosystem, utilizing a dynamic database of dependencies and contributions to Rust libraries assembled by Schueller et al.\cite{schueller2022rust}. Rust is a relatively young but popular and growing programming language, which was recently adopted as the second official language of the Linux kernel project. We choose Rust for several reasons. First, the Rust dependency manager Cargo stores valuable data on the evolution of dependencies between libraries overtime. It also has data on the number of downloads over time, allowing us to test the impact of hypothetical failures on end users. Second, a large majority of Rust libraries are hosted on GitHub or Gitlab, likely because of the language's youth relative to these platforms and its community's strong OSS orientation, allowing us to download nearly all libraries and their complete development histories. Finally, as a growing ecosystem, Rust allows us to track the evolution of systemic risk across its life-course. We note that while other ecosystems may not have the same quality and scope of data, our framework is modular and can be adapted to different datasets.

\subsection{Package metadata, repositories, dependency network}
The core of the dataset is derived from a database dump from Cargo (available on \url{https://crates.io/data-access}, updated daily) containing extensive metadata about packages (called crates in the Rust ecosystem). The dataset includes package names, URLs, versions, dependencies, creation date, and daily downloads. URLs can be linked to valid repository URLs on GitHub and Gitlab, which can be cloned locally. It also provides details on dependencies between packages, enabling the construction of a dependency network at various points in time, which is important as libraries add and remove dependencies on a regular basis. The database discards information about versions by considering only the dependencies of the latest version of a library - recognizing that version conflicts are a major way in which libraries break because of undermaintenance \cite{decan2019empirical,wang2020watchman}.

\subsection{Developer Contributions}

We consider commits as the elemental contributions that developers make to projects. Though we acknowledge that other forms of contributions such as issue reporting represent valuable contributions to OSS projects and ecosystems as a whole \cite{trinkenreich2020hidden}, adjusting to upstream issues, for example, typically requires committing code. To quantify the extent to which individuals contribute to specific packages, commit authors need to be disambiguated because commits themselves are only signed with email addresses. The database we use disambiguates contributor emails to the level of GitHub and Gitlab accounts using the respective APIs \cite{montandon2019identifying,schueller2022rust}. This approach makes more accurate merges than commit email address approaches \cite{fry2020dataset}, which rather have the benefit that they scale to larger datasets. The database we use also labels contributions from bots, using a ground truth dataset  \cite{golzadeh2021ground} and a manual inspection of the most active account, which we exclude from our analyses. 

Finally, we considered two approaches for associating developers to libraries as contributors. Often researchers consider some threshold, for example the smallest set of contributors accounting for at least 80\% of activity in the library \cite{mockus2002two,robles2009evolution}. We rather considered all developers making any contribution to a library in the year preceding a chosen reference time, weighing them by their share of contributions (in number of commits). In our application, we felt this was a more appropriate choice because (relatively) infrequent contributors may be good candidates to step in when additional help is needed, and their work may still constitute a non-negligible share of the global work done on repository. Our code and methods can easily be adapted to consider only core developers according to a threshold of activity, or only contributors with merge rights, or to study activity at different time scale (e.g. a month instead of a year). 

In the analysis carried out in the rest of the paper we fix the scope of the dataset: we consider the dependency network between Rust repos as observed on January 1, 2022. We consider contributions (commits) made to repos from January 1, 2021 to January 1, 2022. We operate in repo space rather than package space, noting that a similar analysis can be carried out in the latter. To emphasize this point of flexibility, we refer to \textit{libraries} rather than repos or packages in the subsequent sections.

\subsection{Data and Code availability}

An anonymized dataset is available on Figshare (\url{https://figshare.com/s/93158d03416765444650}). The underlying code for both data collection and processing is released as an open-source Python library called RepoDepo: \url{https://github.com/wschuell/repodepo}. RepoDepo acts as a wrapper around a database structure in either SQLite or PostgreSQL, with specific adaptors to fill the database from different data sources (CSVs, GitHub/Gitlab REST or GraphQL APIs, and crates.io daily database dumps) and to extract data/apply specific processing. The structure is meant to be modular so that custom functions to collect more data can be added later on. Code to reproduce our analyses is available at \url{https://github.com/wschuell/misteriosse}.

\section{Methods and Analysis}

\subsection{Modeling Library Functionality via Production Functions}
The key insight our paper brings from the complex systems literature is that when systems have rich interdependencies, small changes in seemingly unimportant parts of a system can have an outsized effect on the functioning of the whole. To carry out this kind of analysis in the context of spreading failures in the Rust ecosystem, we will now describe how to quantify a library's functionality in terms of social and technical inputs. 

A library $i$ requires both functioning upstream dependencies and active contributors to continue to function properly. A developer stopping to contribute, or a dependency missing, compromised or exposed to bugs will expose the library itself to an increased risk. We assume that both sources of risk can be combined into one quantity: a probability of \textit{failure} $0\leq F(i) \leq 1$. Risk is minimized when a library has active maintainers and functioning dependencies. Risk is highest when all developers having stopped maintenance work on the library and/or all upstream dependencies have failed. To combine the two sources of risk, adapt the notion of a \textit{production function} from the economics literature \cite{brown1957meaning}. Production functions are used in a variety of contexts to describe how inputs are combined to generate outputs. A traditional example is how capital and labor combine to create goods in an economy. The chosen functional form governs how the inputs interact with one another. For instance, two inputs may substitute for or complement one another. In one extreme case, one or more inputs may be essential to production.

In our case, we argue that maintaining developers and functioning upstream dependencies are both required for a library to continue to work. These two inputs to software maintenance can only substitute for each other in a limited way. This perspective aggregates and necessarily simplifies several sources of risks, but provides a flexible framework to consider the impact of both social and technical vulnerabilities. Specifically we use a Cobb-Douglas style production function \cite{brown1957meaning} $P_i$, which considers the product of the shares of functioning its upstream dependencies ($d_{i}$) and active contributors ($c_{i}$):
$$ F_i = 1 - P_i = 1 - (c_{i}^{1/2}\star d_{i}^{1/2})  $$.

For example, a library or package with one half of its contributors available, and two-thirds of its upstream dependencies functioning, will have a roughly 43\% ($1 - ((1/2)^{1/2}\star (2/3)^{1/2})$) risk exposure, i.e. its chance of failing. We selected the Cobb-Douglas production function to model the spread of risk because it suggests that contributors and upstream health are complements and imperfect substitutes, and that libraries can fail if either is missing. Indeed the Cobb-Douglas functional form is often used to model or estimate the relative contributions of labor and capital to output in a firm or industry \cite{brown1957meaning}.

\begin{figure}
    \centering
    \includegraphics[width=0.4\textwidth]{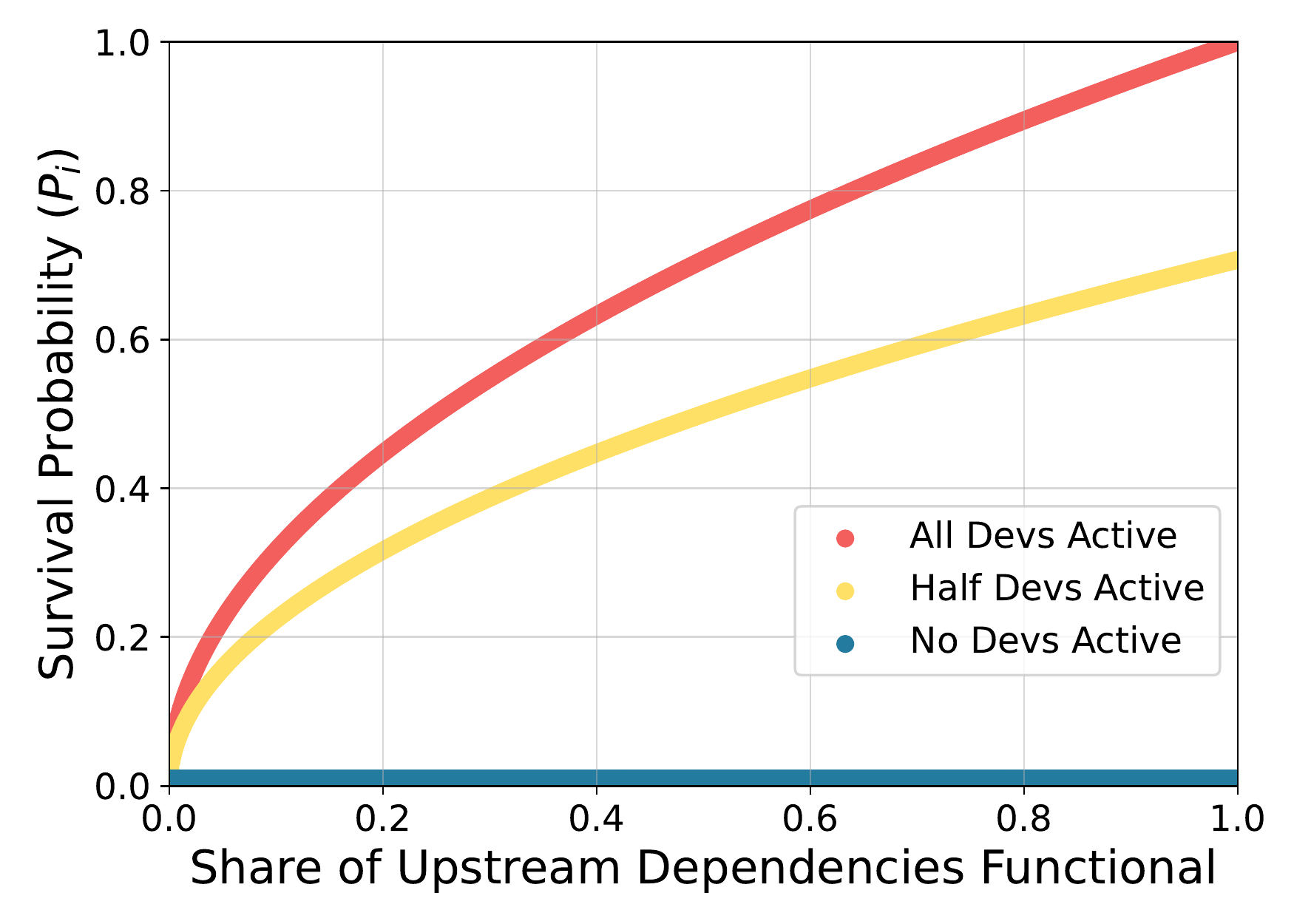}
    \caption{The chance of library failure in terms of the shares of inactive developers and failed upstream dependencies, as quantified using a Cobb-Douglas production function. A library with all of its original developers still active, and 60\% of its direct upstream dependencies functional has a roughly 80\% chance to survive.}
    \label{fig:cobb}
\end{figure}

Other production functions such as the Leontief production function ($1-min(c_{i},d_{i})$), in which inputs cannot be substituted at all, or a linear production function ($1-\frac{c_{i}+d_{i}}{2}$), in which inputs are perfect substitutes, may be more appropriate for the spread of different kinds of issues. In this sense our framework is flexible and can diagnose systemic risk with respect to different kinds of issues. We plot the Cobb-Douglas function predicted chance of library failure in Figure \ref{fig:cobb}. In the following, we study libraries at the level of repositories.

\subsection*{Diffusion of Failures}
In order to model the spread of library failures in the ecosystem, we need represent two kinds of relationships. The first kind consists of maintenance activity by developers in specific libraries. We store this information as a matrix\footnote{We refer to developers and contributors interchangeably, preferring to use the notation $C$ to distinguish their role from the role of technical dependencies.} $C$, in which the entry $C_{i,j}$ counts the number of commits made by developer $i$ to library $j$. We normalize the columns of this matrix and obtain $\hat{C}$, in which the entry $\hat{C}_{i,j}$ can be interpreted as the share of contributions to library $j$ made by developer $i$.

The second kind of relationships within the ecosystem we consider are the dependencies between libraries. We store this information in a matrix $D$. $D$ is a square matrix with rows and columns equal to the number of dependencies. Entry $D_{i,j}$ is equal to 1 if library $j$ depends on library $i$, and is 0 otherwise. Similar to the previous case, we normalize the columns of this matrix to obtain $\hat{D}$, in which the entry $\hat{D}_{i,j}$ can be interpreted as the share of dependency of library $j$ on library $i$.

We now define two vectors that track the state of the system. $S^{C}$ is a vector corresponding to the contributors in the ecosystem. The $i$-th entry of $S^{C}$ is 1 if contributor $i$ is active, otherwise 0. As our analysis deals with the potential consequences of contributors leaving the ecosystem, this vector will be an input to our scenarios. 

The second vector $S^{L}$ tracks the state of each library. That is to say it defines the likelihood that a library will fail, given the status of its upstream dependencies and contributors.

When all libraries are fully functioning, every coordinate of $S^{L}$ is equal to 1. Those coordinates correspond conceptually to $1-F_i$ as defined in the previous subsection, and depending on the imposed conditions -- e.g. some developers missing -- can take values between 0 and 1, 0 being the highest possible level of risk exposure. 

Our scenarios consider what happens to the libraries after a contributor leaves the ecosystem. The departure of a developer triggers potential issues: either directly on those libraries to which she contributes, or indirectly, on those libraries which depend on libraries she maintains. This information is captured by the following self-referential equation:

$$S^{L} = (S^{C^\intercal}\star \hat{C})^{1/2} \odot (S^{L^\intercal}\star \hat{D})^{1/2}.$$

In this equation $\odot$ denotes element-wise matrix multiplication. Likewise, the exponents are to be taken element-wise. $\intercal$ denotes the transpose of the vectors. In plain terms, the left factor corresponds to the effect of absent contributors on the states of the libraries, while the right factor corresponds to the effect of potential malfunctioning upstream dependencies. These effects are combined by the Cobb-Douglas style production function. 

In our application $\hat{C}$, $\hat{D}$, and $S^{C}$ are fixed and we can represent the equation in the following form $S^{L} = f\left(S^{L}\right)$, emphasizing that the library state vector is updating. To find the solution of this equation, we iterate from the initial state in which all libraries are functioning, which we denote $S^{L}_0$:

$$ S^{L}_0 =\mathbf{1} $$
$$ S^{L}_{n+1} = f\left(S^{L}_{n}\right)$$

As we will discuss below, this sequence converges towards a solution in a finite number of steps, which we denote $S^{L} = S^{L}_{Fin}$. Subsequent iterations do not update the vector. 

We now describe the specific analyses we carry out. As discussed, we proceed by initializing all entries of the library state vector $S^{L}$ to 1. We then remove a developer from the system, changing a single entry of the contributor state vector $S^{C}$ from 1 to 0. This allows us to calculate a step of the spread or diffusion of issues. We note that our practical implementation includes two small corrections to the equation to handle edge cases. When a library has no dependencies, its entry on the right factor is set to one at the end of every step in the iteration. Similarly for the left factor for libraries with no default contributors.

In the next step of the spreading, note that the left factor is unchanged - we do not remove additional developers - and that the right factor is simply the result of the calculation carried out in the previous step. We repeat this calculation several times, until the state vector of $S^{L}$ is unchanged. As the dependency network has no cycles (i.e. is a directed acyclic graph or DAG), the process always converges. In practice, this happens in a relatively small number of steps -- approximately 20 in our application -- as the depth of the dependency tree (through which faults spread) is small relative to the size of the system as a whole. 

\begin{figure}
    \centering
    \includegraphics[width=0.5\textwidth]{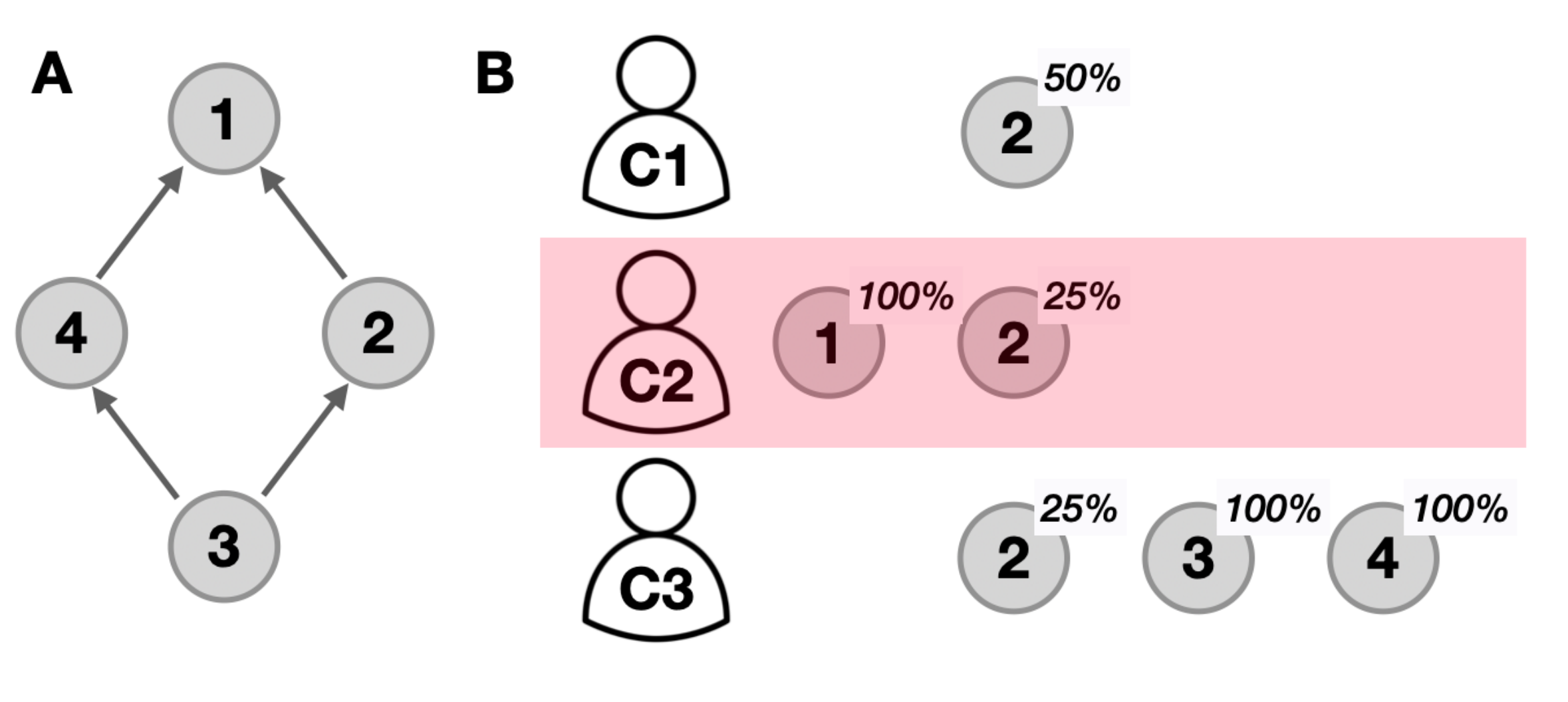}
    \caption{An example ecosystem. A) The network of dependencies between four libraries. B) The three contributors to the libraries. Percentages denote what share of contributions they make to a specific library. For instance, contributor 1 makes 25\% of all contributions to library 2. In our example calculation, we simulate the consequences of the departure of contributor 2 from the system.}
    \label{fig:toy}
\end{figure}

We now carry out one step of an example diffusion on a toy ecosystem. We visualize this example in Figure~\ref{fig:toy}. In this ecosystem there are four libraries maintained by four developers. We define the normalized contributors matrix: 

$$ \hat{C}= \begin{bmatrix}
0 & .5 & 0 & 0 \\
1 & .25 & 0 & 0 \\
0 & .25 & 1 & 1 
\end{bmatrix}  $$

Column 2 indicates that contributor 1 is responsible for half of the contributions to library 2, while contributors 2 and 3 are responsible for one quarter each. Contributor 2 is the sole contributor to library 1. The normalized dependency matrix, on the other hand is a square 4x4 matrix:

$$ \hat{D}= \begin{bmatrix}
0 & 1 & 0 & 1 \\
0 & 0 & .5 & 0 \\
0 & 0 & 0 & 0 \\
0 & 0 & .5 & 0 
\end{bmatrix}  $$

This matrix indicates that library 1 has no upstream dependencies (column 1), while libraries 2 and 4 depends solely on library 1. Library 3 (column 3) depends on both libraries 2 and 4.

We now calculate what happens according to our method if contributor 2 is removed from the system, as indicated in Figure~\ref{fig:toy}. We obtain the following equation:

$$ 
S^{L}_1 = 
(\begin{bmatrix}
1 & 0 & 1 
\end{bmatrix}
\star
\hat{C})^{1/2}
\odot
(
\begin{bmatrix}
1 & 1 & 1 & 1
\end{bmatrix}
\star
\hat{D})
^{1/2}
$$

Carrying out the matrix multiplications and the exponents (element-wise) of the two factors results in:

$$
S^{L}_1 = 
\begin{bmatrix}
0 \\
\sqrt{\frac{3}{4}} \\
1 \\
1 
\end{bmatrix}  
\odot
\begin{bmatrix}
1 \\
1 \\
1 \\
1 
\end{bmatrix}  
$$

As the multiplication is element-wise, the updated vector is simply the first factor:
$$
S^{L}_1 = 
\begin{bmatrix}
0 \\
\sqrt{\frac{3}{4}} \\
1 \\
1 
\end{bmatrix} 
$$

In the next step of the calculation, the left factor would be unchanged, but the right factor would be changed, reflecting the spread of the probability of failure from libraries 1 and 2 (recall that the removed contributor 2 contributed to both). As the reader can verify, the removal of contributor 2 leads to a chain reaction in which the functionality of all libraries in the ecosystem is affected. Specifically the next step of the iteration yields:

$$
S^{L}_2 = 
\begin{bmatrix}
0 \\
0 \\
\sqrt{\dfrac{\sqrt{3}+2}{4}} \\
0
\end{bmatrix}.
$$

The next iteration after that results in a library state vector of all zeros. In other words, $S^{L}_3 = S^{L}_{Fin} = \vec{0}$.

\subsection*{Ranking Contributors, Libraries, and the Ecosystem as a Whole}

With this method to model the spread of failures resulting from the removal of individual contributors, we proceed to test the robustness of the whole Rust ecosystem, aiming to rank contributors and libraries for their systemic importance. To do so we simply repeat the iterated calculation above for every contributor in the ecosystem. That is, we remove each contributor, alone, a single time. Note that more complex removals are possible - the diffusion equation is flexible and can accommodate any valid input contributor state vector. For example, one could remove all contributors supported by a specific corporation or foundation.

For each contributor we remove from the system, our calculations yield a final library state vector $S^{L}_{\text{Fin}}$. These vectors represent the functionality of the libraries following the cascade induced by a specific contributor's departure. We define the risk $R_{i}$ to library $i$ as the difference:

$$R_{i}= 1 - S^{L}_{\text{Fin}}(i)$$.

As not all libraries are equally important to the ecosystem as a whole, we weigh these entries by the share of downloads the specific library has of all total downloads. Specifically the download-weighted risk of a simulated removal to library $i$ is defined as:

$$R^{dl}_{i}= (1 - S^{L}_{\text{Fin}}(i))\cdot \dfrac{dl_{i}}{\sum_{j}dl_{j}},$$

where the denominator in the right factor of the product denotes the sum of downloads of all libraries. Recall that downloads (like commits) are counted only in the year January 1, 2021 to January 1, 2022 for the analysis and results presented in this paper.

Summing the resulting download-weighted risk score over all libraries in the ecosystem yields our final risk score for a specific scenario (defined originally by the input contributor state vector $S^{C}$. It also carries a straightforward interpretation: it describes the average risk exposure of a random individual download of a library in the ecosystem for the given scenario. With this framework in hand we can now define how we rank contributors and libraries, and quantify overall ecosystem's risk.

\paragraph*{Ranking Contributors}
Ranking contributors in terms of their systemic importance in this context is straightforward. Given the removal of contributor $j$, implemented by setting the $j$-th entry of the input contributor state vector $S^{C}$ to 0, the overall \textbf{Contributor Impact} $I_{j}$ is simply the sum of all library download-weighted risk scores:

$$I_{j}=\sum_{i} R^{dl}_{i}.$$

In other words, this measures the average impact that the removal a contributor from the system would have on a random observed download of a library in the Rust ecosystem.

\paragraph*{Overall ecosystem risk}
We derive global measure $G$ of the systemic risk of the ecosystem by summing up all contributor impact scores:

$$G=\sum_{j} I_{j}.$$

This quantity can be used to compare the change in overall risk due to some intervention, as we will implement later in the paper. It also provides a baseline which we use to define library importance.

\paragraph*{Ranking Libraries}
Finally, we also derive a measure ranking the systemic importance of libraries. This is perhaps the most important ranking, as interventions can most easily be made at the level of libraries. Specifically, we  consider how often a library serves as a conduit of spreading failure to downstream dependencies. 

We do this by rerunning the full set of developer removals and failure propagation calculations with each library ``immunized'' to failure, one by one. In terms of our equation, the immunized library's entry in the library state vector $S^{L}$ is hard-coded to 1. This counterfactual allows us to quantify how a library amplifies or \textit{transmits} issues through the software dependency network. The change in the overall ecosystem risk score $G$ when a library is protected in this way serves as a quantification of its contribution to overall risk. We define this measure, which we call the \textbf{Risk Transmission Score} of a library $i$, as follows:

$$ RTS_{i} = G - G_{i},$$

where $G_{i}$ denotes the ecosystem risk score calculated when library $i$ is immune to failure. 

\begin{figure}
    \centering
    \includegraphics[width=0.4\textwidth]{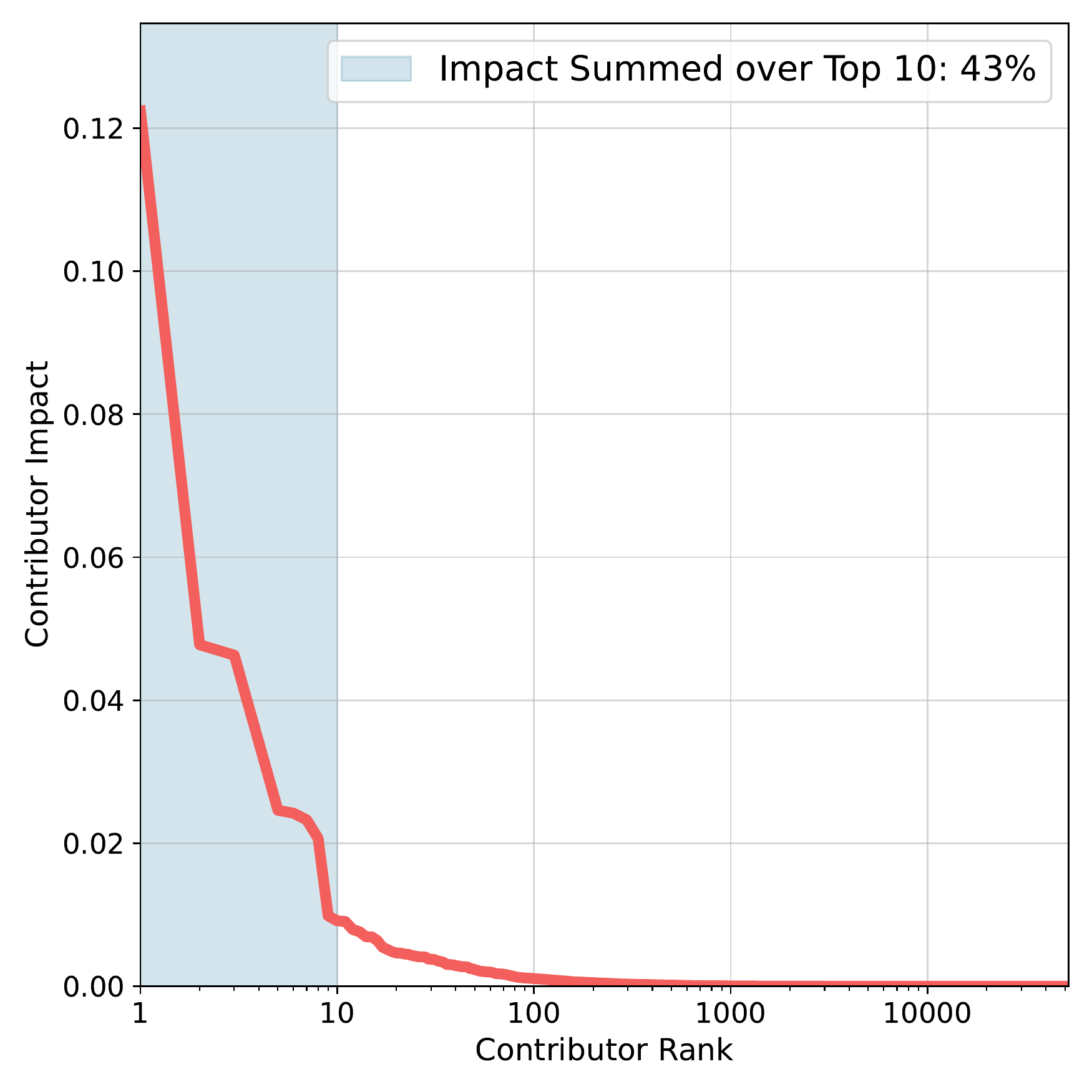}
    \caption{The rank-ordered contributions of Rust contributors to systemic risk, on a logarithmic scale. The top 10 developers account for 43\% of systemic risk.}
    \label{fig:devranks}
\end{figure}

\subsection*{Results}
We plot the rank-ordered distribution of Contributor Impact in Figure~\ref{fig:devranks}. We observe a remarkable concentration of systemic risk: with the top 10 contributors accounting for over a 40\% of the risk observed in our analyses. We observe a similar, though slightly less concentrated distribution when we consider the importance of different libraries in terms of their Risk Transmission Score. This recalls our motivating example from earlier in the paper: individual developers are playing an important role in multiple important libraries.

\begin{figure}
    \centering
    \includegraphics[width=0.4\textwidth]{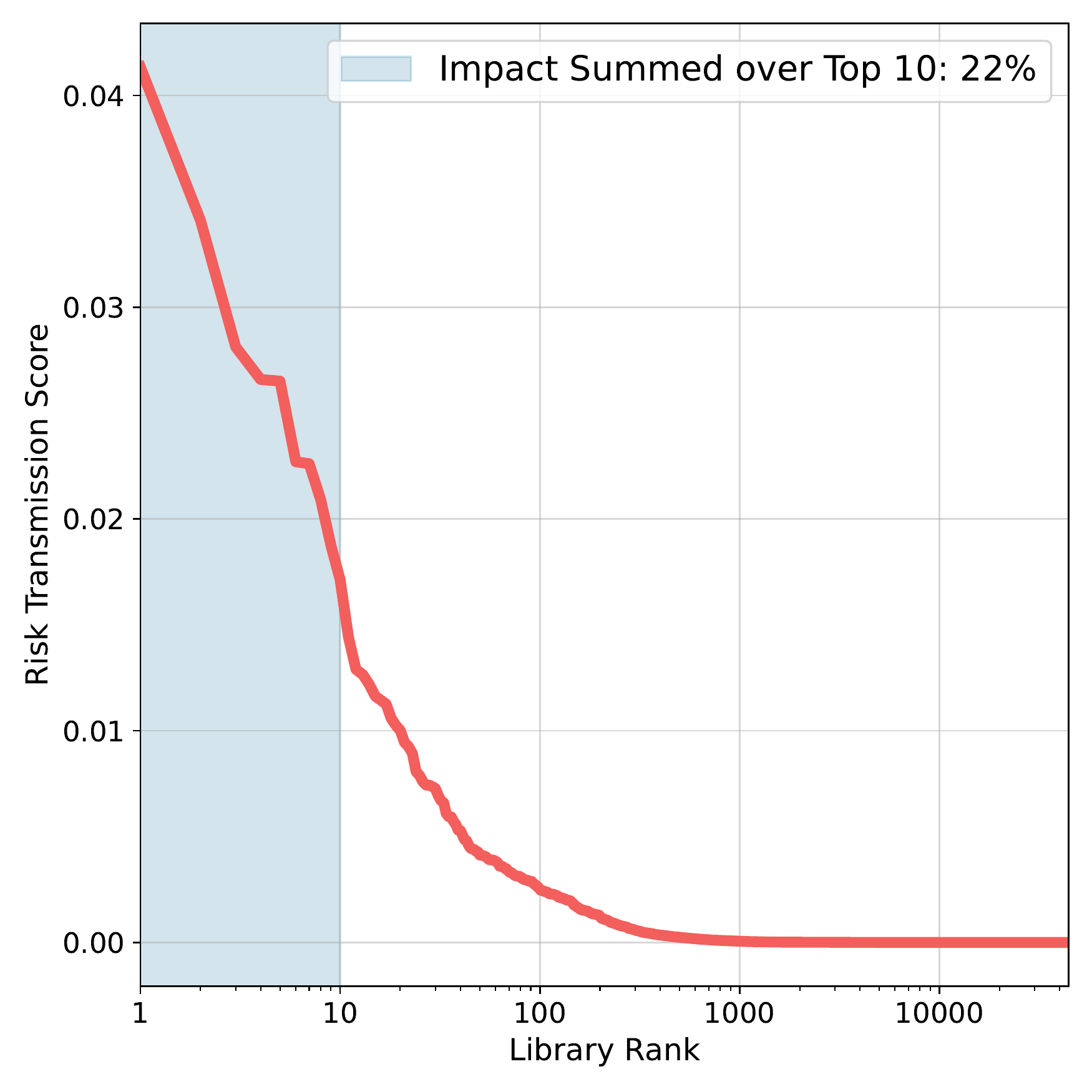}
    \caption{The Risk Transmission Rank (RTR) importance of Rust libraries, on a logarithmic scale. We quantify a library's importance by rerunning our failure cascade model with that library immunized against failure, and then comparing the overall outcome against the general case when the library can fail. The top 10 libraries account for 22\% of the total observed differences across all libraries.}
    \label{fig:libranks}
\end{figure}

\begin{figure}
    \centering
    \includegraphics[width=0.5\textwidth]{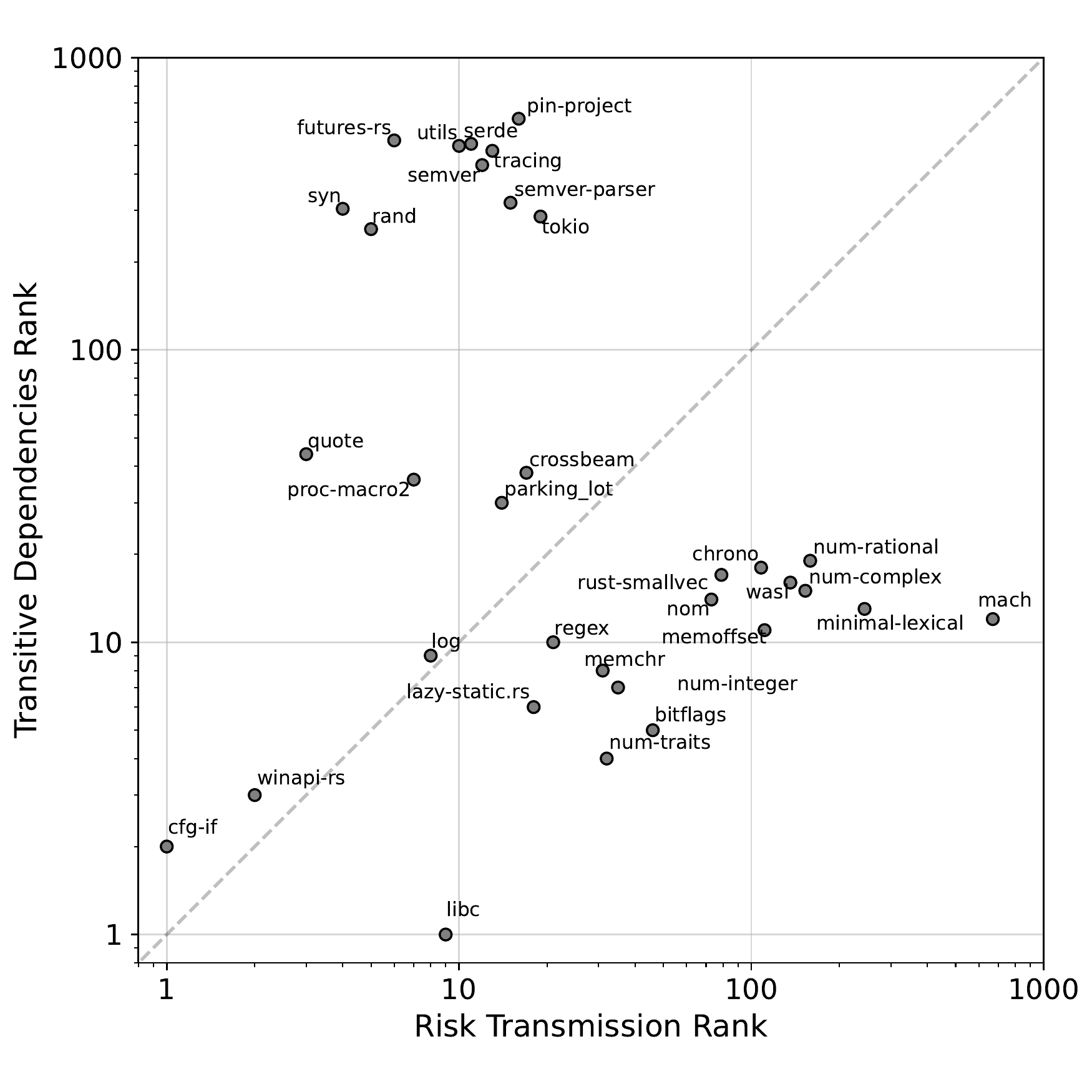}
    \caption{Libraries of the Rust ecosytem in the top 20 according to either their Risk Transmission Rank (RTR) or the count of (downstream) transitive dependencies. We observe several libraries in the top 10 of the RTR, hence likely of systemic importance, that are not even in the top 100 according to the number of transitive dependencies they have.}
    \label{fig:rtr_vs_trans}
\end{figure}

To compare our method of ranking important libraries with alternative measures, for example by the count of their transitive dependencies, we zoom in on the very top ranked libraries. We plot libraries ranked among the top 20 according to Risk Transmission Rank \textit{or} by the count of their transitive dependencies in Figure~\ref{fig:rtr_vs_trans}. Our method suggests that libraries above the diagonal are more important than a count of their downstream dependencies suggests. Our method highlights several libraries (``rand'' and ``syn'', among others) that are among the top 10 libraries according to Risk Transmission Rank, but do not break into the top 100 libraries by number of transitive dependencies. These libraries have a prominent position in the network topology that amplify their contribution to systemic risk.

\section{Interventions}

While it is valuable to highlight vulnerabilities in a system, our methodology can be used to suggest how to intervene in the system to improve its resilience by allocating scarce development resources. In particular, our ranking of libraries in the Rust ecosystem can be used to allocate support. For instance, a foundation or firm may have funds to sponsor development or maintenance on a specific library. Though in reality developer resources are not fungible and cannot be allocated to any library in arbitrary amounts, a prioritization in terms of risk remains useful. Thus in this section we describe an intervention in terms of development time contributed to a fixed number of libraries. We compare the impact on systemic risk of various allocation strategies based on rankings of libraries, including our Risk Transmission Rank score. We first describe alternative rankings, then describe how we implement the interventions, and finally report results.

\subsection{Rankings}
As our aim is to compare reasonable strategies to allocate development resources across libraries, we consider several perspectives on what makes a library important or well-known. We consider a library's place in the dependency network, its overall use, its age, its popularity, and its systemic importance as quantified by our Risk Transmission Rank. We also test random allocations as a benchmark. We summarize these rankings in Table~\ref{tab:ranks}, and describe them below.

\begin{table}[!t]
\ra{1.1}
\setlength\tabcolsep{14pt}
\begin{tabular}{lp{0.22\textwidth}}
\toprule
\setlength\tabcolsep{2.5pt}
Name & Description\\
\midrule
Transitive Dependencies & Count of upstream dependencies\\
Downloads & Number of downloads on Cargo \\
Age & Time of first appearance on Cargo \\
Stars & Number of stars on GitHub \\
Random & Random allocation (baseline)  \\
\textbf{Risk Transmission Rank} & \textbf{Spreading-based measure}  \\
\bottomrule
\\
\end{tabular} 
    \smallskip
  \caption{Ways to rank libraries in the Rust ecosystem to allocate developer resources in an intervention.} 
  \label{tab:ranks}
\end{table}

In a technical sense, a library is important in an ecosystem if many other libraries depend on it. These dependencies can be direct or indirect (sometimes called transitive). We therefore use the count of \textit{transitive dependencies} as one ranking of libraries \cite{decan2019empirical}. A simpler way to rank libraries is by their age, arguing that older (active) libraries are more likely to be important.

In practice, libraries are often ranked by their use and popularity in the broader community. We measure use via the count of \textit{downloads} of a Rust library, as tracked by Cargo. Popularity or social visibility is measured by counting the number of \textit{stars} a library has received on GitHub \cite{borges2018s}. Both factors are thought to play an important role in the success of open source software. A large active user base of a specific library provides a kind of defense against errors as suggested by Raymond's notion that ``with many eyes all bugs are shallow'' \cite{raymond1999cathedral}. The launch of GitHub Sponsors and growing adoption of crowdfunding to support OSS maintainers suggests that highly visible libraries will be even more likely receive resources \cite{overney2020not}. However we know from examples that not all systemically importance packages are highly visible. Anecdotally, OpenSSL was taken for granted before the discovery of the Heartbleed vulnerability.

\subsection{Intervention Design and Quantifying Impact}

While software developers and software development time is not a fungible resource, we make the simplifying assumption that a donor can contribute to a library by adding a single developer. Developers added this way make a uniform weekly contribution of commits, which we fix at 5/7 times the number of days across which we analyze the system (in our case 365, covering all of 2021). We denote this contribution factor (5/7 times 365) by $e$. This simplification suggests that allocated developers make roughly one contribution per weekday. In our framework this level of contribution can be tuned. In any case, we add a single developer to the top $K$ libraries according to each ranking for a range of values between 1 and 1000. 

After allocation we rerun our framework - one by one we remove each developer in the system and calculate the resulting cascades. We calculate $G$, the overall systemic risk of the ecosystem in each of these scenarios, comparing it the original estimate derived in the previous section. As more contributors are added, the overall systemic risk falls. Comparing which ranking method decreases risk how much and how quickly across a range of values will tell us about the effectiveness of interventions.

There is one modification we need to make to the methodology in this case. Specifically, if we are adding contributions to a library as part of an intervention, we should not simulate what happens if these new resources are withdrawn. Rather, we represent these extra contributions as a kind of \textit{surplus} or \textit{overproduction} (c.f. \cite{champion2021underproduction}) that can be used to absorb shocks. Specifically, we add a term $\textbf{X}$ to our self-referencing equation for the library state vector:

$$S^{L} = (S^{C^\intercal}\star \hat{C})^{1/2} \odot (S^{L^\intercal}\star \hat{D} + \textbf{X})^{1/2}.$$

Here $\textbf{X}$ is a vector corresponding to the libraries of the ecosystem. An entry $i$ is equal to $e/N$, where $N$ is the total number of commits to library $i$, if library $i$ is allocated a developer, 0 otherwise. Recall that $e$ captures the contributions of these allocated developers, estimated at a rate of one commit per week day. This correction insures that additional resources allocated by the intervention can only help a library. In each round of the calculation, we cap entries of the library state vector $S^{L}$ at 1 to insure convergence.

\subsection{Intervention Results}
\begin{figure}
    \centering
    \includegraphics[width=0.45\textwidth]{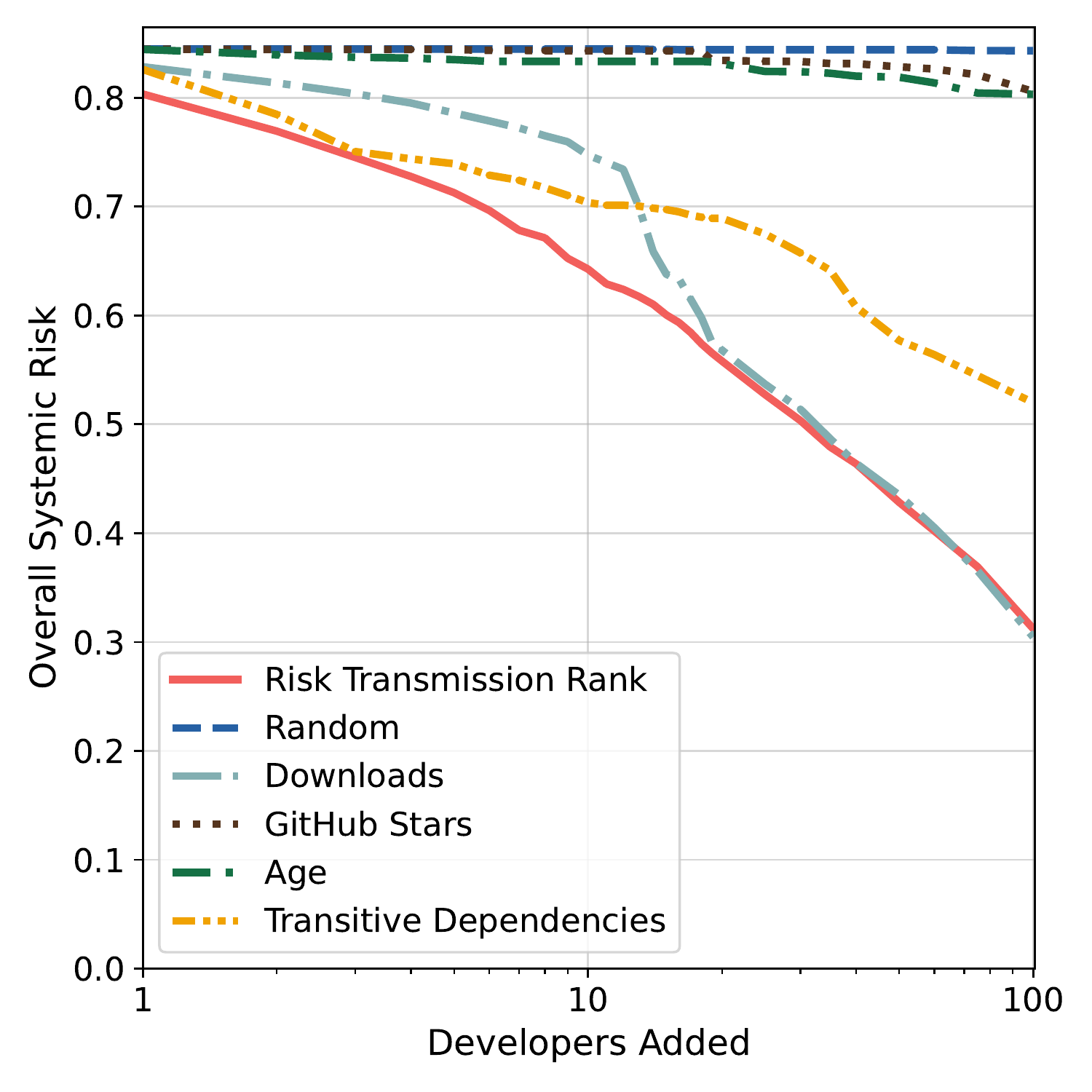}
    \caption{The change in overall systemic risk as a function of developers added by various library ranking heuristics, ranging from adding one contributor to 100 (log scale).}
    \label{fig:adding100_devs}
\end{figure}

\begin{figure}
    \centering
    \includegraphics[width=0.45\textwidth]{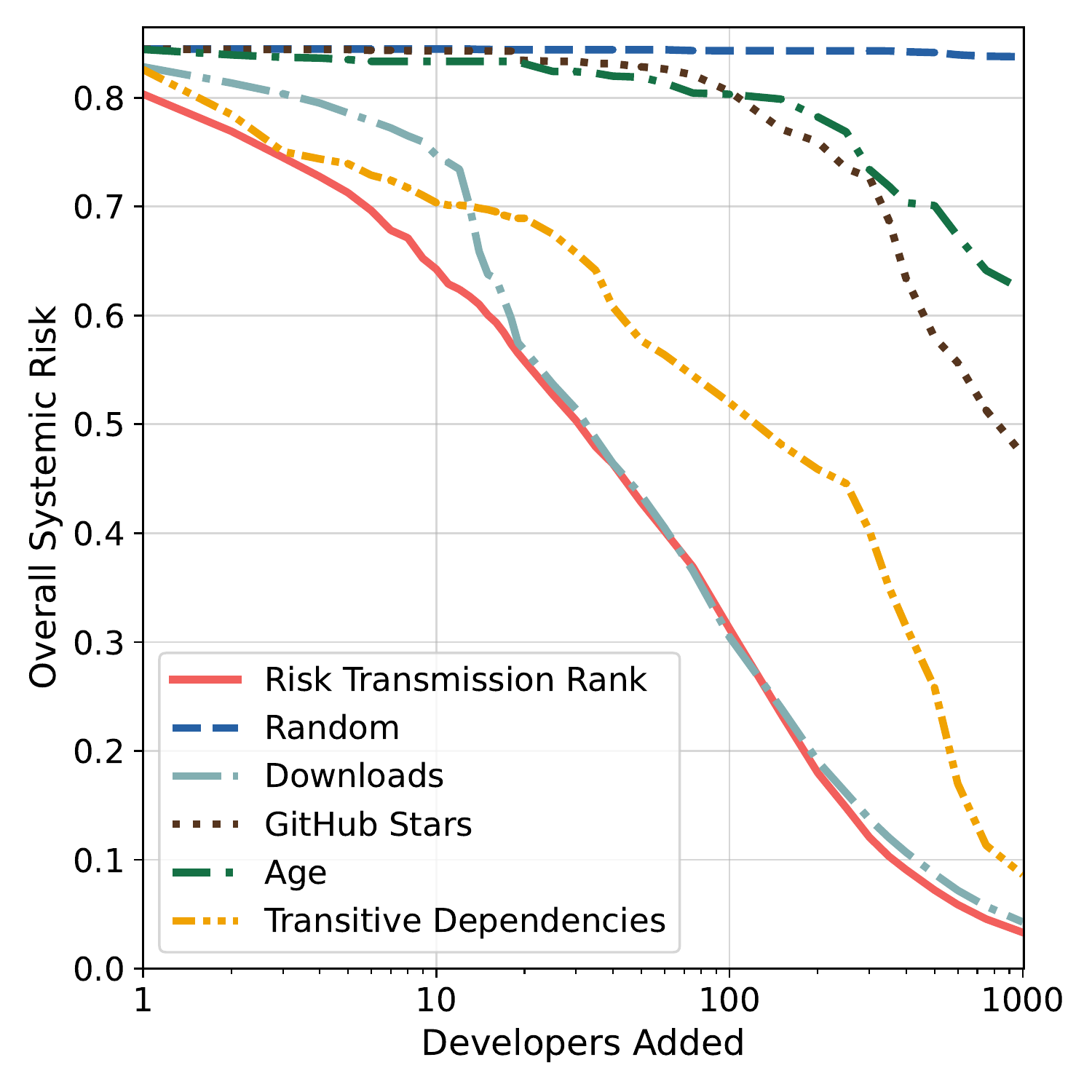}
    \caption{Extending the previous figure to scenarios adding between 100 and 1000 developers to the ecosystem.}
    \label{fig:adding1000_devs}
\end{figure}

We visualize the improvement in systemic risk as a function of developers added according to the different ranking heuristics in Figures \ref{fig:adding100_devs} and \ref{fig:adding1000_devs}. We observe a sharp decrease in overall risk using the Risk Transmission Rank, the number of downloads, or the number of transitive dependencies to rank libraries. Using GitHub stars, library age, or a random ranking to allocate development resources are significantly less effective strategies. Given that GitHub stars are a major source of visibility in the OSS community \cite{borges2018s}, we suggest that equating systemic importance with stars may be leading to systemic mis-allocation of attention and help.

\begin{table*}[t]
\ra{1.1}
\begin{tabular}{lcccccccccc}
\toprule
Developers Added: &            1    &  2    &  5    &   10   &   20   &   50   &   100  &   250  &   500  &   1000 \\
\midrule
Intervention      &                 &       &       &        &        &        &        &        &        &        \\
Risk Transmission Rank &   2.4\% &  4.7\% &  9.5\% &  14.8\% &  21.9\% &  34.3\% &  45.3\% &  62.9\% &  75.4\% &  84.9\% \\
Downloads    &    1.0\% &  1.9\% &  4.0\% &   6.5\% &  14.5\% &  30.8\% &  43.7\% &  61.8\% &  74.0\% &  83.4\% \\
Transitive Dependencies &            1.1\% &  2.9\% &  7.7\% &  11.2\% &  14.4\% &  20.7\% &  28.0\% &  37.5\% &  48.6\% &  66.3\% \\
GitHub Stars     &            0.0\% &  0.0\% &  0.0\% &   0.1\% &   0.2\% &   1.0\% &   2.0\% &   6.3\% &  14.0\% &  26.2\% \\
Age             &     0.0\% &  0.1\% &  0.5\% &   0.8\% &   1.0\% &   2.1\% &   3.6\% &   5.7\% &   9.4\% &  15.7\% \\
Random    &    0.0\% &  0.0\% &  0.0\% &   0.0\% &   0.0\% &   0.1\% &   0.1\% &   0.1\% &   0.2\% &   0.5\% \\

\bottomrule
\end{tabular}
\smallskip
\caption{Cumulative reduction of systemic risk when adding developers according to different intervention strategies, relative to the baseline systemic risk. Whether adding just a few or many developers, adding them to high RTR ranked libraries yields the greatest decrease in systemic risk.}
\label{tab:intervention_results}
\end{table*}

To quantify the relative performance of these rankings, we calculate the area below the horizontal line at the baseline systemic risk level and above the curve for each heuristic up to different numbers of developers added, and normalize by the total area under the baseline. We report these values in Table \ref{tab:intervention_results}. These results verify the patterns we observe in the figures: allocating developers by RTR, downloads, and transitive dependencies are better strategies than allocating them by age or visibility (i.e. stars on GitHub). Though RTR performs best across the entire range of the intervention, it is still interesting to note that ranking libraries by transitive dependencies and downloads are also relatively strong heuristics. As these require less data and calculation, they may be good strategies in other ecosystems with less available data.

\section{Discussion}

OSS ecosystems tend to evolve, like many complex systems, towards efficiency. If there is a library that does something well, it can quickly become widely used. However, this drive towards efficiency may increase systemic risks. Studies of risk in ecosystems that focus on the structure of technical dependencies overlook the potential synchronized risks coming from developers active in multiple libraries. In this work we presented a framework to quantify these risks. Our method highlights individual libraries which are worthy of more attention and support. As the most central and important developers in OSS ecosystems are under increasing pressure and stress \cite{eghbal2020working}, measures of ecosystem health need to consider the interaction of these social aspects with their technical structure \cite{constantinou2017socio}. Within individual libraries or projects, the importance of socio-technical congruence - that is the coordination between people working on interdependent modules - is well known \cite{cataldo2008socio}; our paper suggests how such congruence at the ecosystem level could provide warnings and help maintainers of downstream libraries anticipate upstream issues. More generally, our work contributes to a growing literature on the sustainability and resilience of key software supply chains \cite{zimmermann2019small,ohm2020backstabber,lamb2021reproducible}.

Our framework has several potential extensions. The most natural one is to generalize the approach to other ecosystems \cite{franco2017open}, as done for the analysis of dependency networks in Decan et al. \cite{decan2019empirical}. Given that dependency networks are known to be growing in several ecosystems, and that the truck factor of key libraries in multiple ecosystems are known to be low, we are confident that the application of our framework to other systems will yield a similarly useful diagnosis of points of vulnerability. We chose to focus on Rust given that temporal data on downloads of individuals is available, and that a significant majority of libraries are available on GitHub. This suggests a threat to generalizability: many package ecosystems do not disclose data on downloads. Here we note that our approach can be adapted to such cases by simply equalizing download weights across all libraries, or using some other measure of or proxy for use as a substitute. It is also possible to define dependencies, and subsequently ecosystems, via references in code \cite{blincoe2015ecosystems}. The general idea of our work can be applied to other contexts in the software industry in which people reuse and share code, for instance the case of Docker images \cite{cito2017empirical}.

Another potential threat to validity and point to improve is that we only use commits to determine contributors. While commits are needed to adapt to issues and bugs from upstream dependencies, they are not the only way to help an open source project, and a more complete accounting of contributions would include work on issues and community management \cite{trinkenreich2020hidden}. Even among commit activity, future work could go into greater depth by inferring which developers are maintaining which parts of a library \cite{gote2021analysing}. However, in general it is difficult to estimate which contributors are most essential to open source projects \cite{casari2021open}, and perhaps even more difficult to know which developers can take over for particular colleagues. We can also test what happens to libraries downstream from abandoned libraries \cite{avelino2019abandonment}, to validate whether focusing on commits is appropriate.

Another simplifying assumption of our work is that the  libraries we observe are functioning and maintained, however many developers they have. This is often false, as shown by Champion and Mako Hill in their recent work on the Debian package ecosystem \cite{champion2021underproduction}. They introduce the the notion of underproduction and quantify it by tracking issue survival rates. One could extend our model by adapting this measure.

Our framework can also be adapted to analyze the consequences of multiple developers leaving an ecosystem at the same time. This is not a hypothetical scenario - the increased reliance on centralized sponsors of libraries and even ecosystems presents another kind of risk. For example, many of the core developers of the Rust ecosystem were employed by Mozilla. In a round of layoffs in the summer of 2020, many of these developers lost their jobs at the same time. While Rust seems to have weathered this storm, it demonstrates that often multiple developers leave a system at around the same time. 

In general, people leave OSS projects for reasons that may be correlated within an ecosystem, such as the end of funding of a university project or changing workplaces \cite{miller2019people}. Luis Villa of Tidelift, a firm that helps users and firms support OSS with financial contributions, suggests that we should perhaps rather talk of a ``boss factor'' than truck factor.\footnote{See: https://blog.tidelift.com/bus-factor-boss-factor-and-the-economics-of-disappearing-maintainers}. Indeed, many of the most widely used and influential OSS projects are maintained by companies and paid individuals \cite{germonprez2019rising}. In this way, OSS ecosystems can be thought of as a co-production of volunteers and companies \cite{o2021open}. Policymakers seeking to promote the use of OSS should consider these aspects of sustainability \cite{blind2021impact}. 

Our work provides additional motivation for getting more people involved in OSS. Mentorship of new contributors has been shown to be a key determinant of people becoming active participants in ecosystems \cite{steinmacher2021being}. At the same time, our work shows indirectly how social barriers to participation  \cite{steinmacher2015social} make software ecosystems more brittle in the long run. More work is needed to understand how disparities in participation in open source, for example owing to gender \cite{terrell2017gender,vasilescu2014gender} or geography \cite{takhteyev2012coding,braesemann2019global,wachs2022geography}, block us from realizing more stable systems.

\section*{Acknowledgements}
The authors thank Christian Diem, Tobias Reisch, Hannah Schuster, Balint Daroczy, Aleksandra Urman, Rositsa Ivanova, and participants of seminars at the Complexity Science Hub Vienna and WU Wien for valuable feedback.

\bibliographystyle{IEEEtran}
\bibliography{IEEEabrv,bibliography}

\end{document}